\documentclass[10pt,a4paper,twoside]{aa}
\usepackage{graphicx}
\usepackage{natbib}
\bibpunct{(}{)}{;}{a}{}{,}
\usepackage{latexsym}
\usepackage{amsmath}
\usepackage{amssymb}
\usepackage{amstext}
\usepackage{color}
\usepackage{psfrag}

\def\cm3{cm$^{-3}$}

\def\12{$^{12}$CO}

\begin{document}
\title {New VLA observations of the SNR \object{Puppis A}: \\
the radio properties and the correlation with the X-ray emission} 

\author{
G. Castelletti\inst{1}
\thanks{ Post-Doc Fellow of CONICET, Argentina}, G. Dubner\inst{1}
\thanks{ Member of the Carrera del Investigador Cient\'\i fico of CONICET, 
Argentina},
K. Golap\inst{2}, \and W.~M. Goss\inst{2}
}

\offprints{G. Castelletti}
\institute {Instituto de Astronom\'{\i}a y  F\'{\i}sica del Espacio (IAFE),
CC 67, Suc. 28, 1428 Buenos Aires, Argentina\\
             \email{gcastell@iafe.uba.ar}
\and
 National Radio Astronomy Observatory (NRAO), New Mexico, USA} 
   \date{Received 2006 Febrary 22; Accepted 2006 July 6}


  \abstract
 {High-resolution, high-sensitivity multifrequency radio 
images of
supernova remnants (SNRs) are essential in advancing the understanding
of both the global SNR dynamics and particle acceleration
mechanisms.}
   {In this paper we report on a new study of the SNR \object{Puppis A} based
on VLA observations at 1425 MHz; the improvement represents a
factor of two in angular resolution and almost ten times in  sensitivity
compared to the best previous image of \object{Puppis A}.
This new image is used to compare
with re-processed 327 MHz data and \it ROSAT \rm and \it Chandra \rm
images to investigate morphological and spectral characteristics.}
   { The observations were carried out with the VLA in the 
DnC and CnB configurations in 2004. 
After combining
with single-dish data from Parkes, an angular resolution of
34$^{\prime\prime} \times 16^{\prime\prime}$ and an rms noise of 0.5
mJy~beam$^{-1}$, were achieved. Archival VLA data at 327 MHz were also
processed. 
The spectral index distribution was then determined by a 
direct comparison of the homogenized data at both 327 and 1425 MHz. In 
addition, to identify different spectral components, tomographic spectral 
analysis was performed.}
   {The new 1425 MHz radio image reveals a highly structured border
encircling a diffuse, featureless interior. 
In particular, the
northern half of \object{Puppis A} displays a complex structure along 
the
periphery, consisting of short arcs resembling ``wave-like'' 
features. 
These are oriented essentially
perpendicular to the shock front on the NE side, but are tangential to the
shock on the NW side. A remarkable
correspondence between such ``wave-like'' features and spectral changes
is observed. On the other hand, the brightest radio features
(located to the E of the SNR and also detected in X-rays)
have no counterpart in the spectral index
distribution. Based on a uniform compilation of
integrated flux densities 
between 19 and 8400 MHz, 
a global spectral index $\alpha = -0.52
\pm 0.03$ (S $\propto \nu^\alpha$) has been determined. 
The new 1425 MHz image of 
\object{Puppis A}
was compared with the \it ROSAT \rm X-ray image of the entire SNR and
with the resolved arcsec \it Chandra \rm image of the bright E
region. There is good overall agreement between the radio
and high resolution X-ray images. At the arcsec level, 
however, the 
agreement
of the radio and X-ray images is less striking.
A noticeable similarity is found between \object{Puppis A}, with its E and W
extensions, and analogous morphological features observed in the SNR 
\object{W50}. This suggests that \object{Puppis A} could be another 
case of a SNR shaped by the action of energetic jets.
}
   {}

       \keywords {ISM: individual objects: \object{Puppis A}-ISM: 
       supernova 
remnants-radio continuum: ISM}

\titlerunning{New VLA images of the SNR Puppis A}
\authorrunning{\textsc{Castelletti et al.}}

\maketitle

\section{Introduction}
Sensitive, high-resolution studies of the radio emitting plasma
associated with supernova remnants (SNRs) have proven to be an 
excellent tool for  
recognizing morphological characteristics, delineating the 
location of 
contact discontinuities in the fluid, and spatially resolving 
filamentary
emission that may indicate expansion into non-uniform environments in
the interstellar medium (ISM).
These high quality radio images are also helpful in addressing questions 
concerning particle acceleration mechanisms 
based on determinations of the spatial variations in the radio
spectral index. 
Moreover, the detailed combination of radio and X-ray
data not only complements the analysis for particle acceleration, but
is also a valuable tool for studying the interaction of the blast 
wave with inhomogeneities in the surrounding ISM.

\citet{dub91} have presented an earlier VLA
image of the SNR Puppis A at 1515 MHz based on observations from a
combination of 26 different
pointings observed with the VLA\footnote{The Very Large Array of the National
Radio Astronomy Observatory is a facility of the National Science Foundation
operated under cooperative agreement by Associated Universities, Inc.} in the
DnC array. The angular resolution of
this image is 77$^{\prime\prime}$ $\times$ 43$^{\prime\prime}$ with a noise of
$\sim\,$3 mJy~beam$^{-1}$. This image shows some agreement between the bright
radio features and the X-ray emission detected by \it Einstein \rm 
X-ray 
Observatory.
However, the angular resolution and sensitivity of this previous VLA
image is not high
enough to allow a detailed comparison with improved X-ray images, 
e.g. that of \citet{pet96} obtained using \it ROSAT\rm, 
or the 
detailed high resolution
\it Chandra \rm study of the eastern region carried out by \citet{hwa05}.
Because of advances in both instrumentation
and imaging algorithms for the production of sensitive
radio mosaics of large regions, a new set of 20 cm radio observations 
have been carried out using the VLA.

In this paper, we present a new 1425 MHz image of \object{Puppis A} generated 
from VLA DnC and CnB configuration data sets obtained in 2003 and 
2004. 
The improved sensitivity
and resolution have led to the discovery of new
morphological features near the outer shock front and 
at the E
side of the SNR shell. From the combination of the radio observations
at 1425 MHz with re-processed data at 327 MHz from 
the VLA archives, 
we have investigated the spatial spectral index distribution across the 
remnant. 
Based on the 1425 MHz image, we have also performed a 
detailed study of the correlation between radio and X-ray emission, 
focusing special attention on the complex region to the E.

\subsection{General multiwavelength properties of the SNR \object{Puppis A}}
\object{Puppis A} is the remnant of a galactic supernova for which 
it is
possible to observe ejecta from the original explosion that
occurred $\sim\,$3700 yr ago \citep{win88}. In addition,  
signatures of the interaction with the
circumstellar/interstellar material can be observed.
Based on an analysis of both X-ray and optical spectra, 
\object{Puppis A} can be classified as belonging to the small class of 
galactic oxygen-rich SNRs
\citep*{can81,win85} in which
fast moving filaments evidence nuclear processes that occurred in the
progenitor star.

The radio emission of \object{Puppis A} is of synchrotron origin and 
takes the form of a highly distorted clumpy shell, with the
brightest section along the E border.
Optical emission is present as a collection of
compact condensations rather than delicate filaments, and shows 
little relation to the radio features \citep{mil83}.

Previous multiwavelength studies of the surroundings of \object{Puppis A} 
suggested  
the existence of dense interstellar clouds along the 
E-NE border of the SNR \citep{pet82,tes87,are90,are91}. 
This was  
later supported by atomic and molecular studies of the surrounding 
interstellar medium; HI and $^{\mathrm{12}}$CO J:1-0 line
observations have revealed the existence of a chain of interstellar clouds 
concentrated along the NE periphery of the remnant \citep{dub88,rey95}.  
From these studies, a kinematical distance of 2.2 $\pm$ 0.3 kpc 
has been derived 
for \object{Puppis A}. 

The X-ray remnant consists of non-uniform emission that includes 
both extended 
features and knots \citep{pet82}, the most conspicuous of which is 
the bright knot located at the E limb 
(called by those authors ``the bright eastern knot'', 
or ``BEK''). 
This feature coincides 
with an ``indentation'' in the radio morphology,  suggesting an 
encounter between the SN shock and an external inhomogeneity. 
Based on an OH 1667 
MHz line study, \citet{woe00} 
suggested that the 
shock front expanding to the E has completely engulfed a cloud
giving rise to the X-ray knot, leaving no observable unshocked 
remainder
of the cloud.
Recently, \citet{hwa05} presented \it Chandra \rm X-ray images and 
spectral analysis of the region around the BEK.
These images reveal details of substructure in this part
of \object{Puppis A}. The spectral analysis shows a temperature gradient 
rising behind
the shock front, suggesting 
possible heating by reflected shocks due to the interaction with denser gas.

An unresolved X-ray source has been detected near the centre 
of \object{Puppis A}
\citep[\object{RX J0822$-$4300}, ][ and references therein]{pav99}. 
This source may 
well be the compact stellar remnant formed in the SN event.
However, no associated pulsar 
wind nebula has been detected \citep{gae00}. 
\section{Observations}
\subsection{Re-processing of radio continuum data at 327 MHz}
327 MHz data acquired with the VLA in the CnB configuration
in 1988 \citep{dub91} were re-processed. 
The data were flagged and calibrated using the AIPS software package.
Further processing was done in AIPS++.
The W-Projection imaging algorithm \citep{cor03}
was used to correct for the non-coplanarity of the VLA array at this
frequency \citep{cor92}. We made use of multiscale clean to
deconvolve the image. The deconvolution was done in iterations, with
the visibility data being self-calibrated in phase after each 
iteration.

To recover the very low spatial frequencies that the VLA 
observations miss, we made use of single-dish   
data extracted from the 408 MHz All-Sky Continuum Survey 
\citep{has81}.
We scaled the flux density of the single dish image with a spectral index 
$\alpha=-0.6$, which we derived from 
comparing the flux densities at 86 and 408 MHz \citep{mil93}. 
The resultant single-dish  
image was combined with our interferometric image.  The
combination is based on a method referred to as ``feathering''. This
method involves Fourier transformation of both the single dish and 
interferometric data onto identical grids. The single-dish Fourier 
data is multiplied by
the ratio of the Point Spread Function (PSF) volumes of the
interferometric image to that of the single dish image. This scaled
data is added to the Fourier interferometric data multiplied by a
feathering function. The feathering function used in AIPS++ is
1-FT(PB$_{\mathrm{dish}}$), where FT(PB$_{\mathrm{dish}}$)  
denotes the Fourier transform of the primary beam
of the single-dish. This combined data is then Fourier
transformed back to the image domain.

Although the angular resolution of the final image at 327 MHz is
90$^{\prime\prime}$ $\times$ 45$^{\prime\prime}$ at a
position angle of 164${\degr}$,
and
the rms noise level is 10 mJy~beam$^{-1}$, comparable to the earlier 
results of \citet{dub91}, the newly processed 327 MHz image 
gives a more reliable representation of \object{Puppis A} at all spatial 
frequencies.
\subsection{New radio continuum observations at 1425 MHz}
We observed \object{Puppis A} at 1425 MHz using the VLA in the DnC and CnB
configurations. The DnC data were obtained on 2003 January 21, 
22, and 23, while the CnB observations were carried out on
2004 February 4, 5, 7, and 15. The flux density calibrators used were 
\object{3C\,147}  
and \object{3C\,286} for the DnC observations and 
\object{3C\,286} for the CnB 
observations. 
The phase and amplitude calibrator used for all the data observed was the 
source \object{J0828$-$375}.
Due to the large size of this SNR (about 55$^{\prime}$ in diameter) we used a
mosaicking technique with 39 different pointings with a spacing of half of 
the primary beam width.
Each sub-field was observed in two 50 MHz bands centered near 1465 
and 1385 MHz in continuum mode. 
The observations were made in cycles with 28 s in
each pointing of the mosaic. This observing cycle was repeated 9 
times in the DnC array and 12 times in the CnB configuration.

The data were reduced using the AIPS++ software package. The data
from each day were calibrated, imaged and self-calibrated
separately to ensure that there were no
day-to-day amplitude discrepancies. 
All the calibrated data sets were then 
concatenated into a single visibility dataset. We imaged and 
self-calibrated 
the combined visibility data in an
iterative fashion. The imaging was done using the standard mosaicking
technique that jointly deconvolves all the fields, taking into account the 
primary beam effect at each pointing. The multiscale clean method was
applied to deconvolve the image. We used
uniformly weighted visibility data for the high resolution image. 
For the image that is used to
compare with the 327 MHz data, the visibility data were tapered in such
a fashion that the resulting restoring beam is identical to that of 
the 327 MHz image.

To recover the large scale structure missed by the interferometer, we used an 
image from the Parkes Southern Galactic Plane Survey \citep{mcc01}.
The single-dish and the interferometric mosaic images were combined
using the same feathering technique described above.

The uniform weighting image that was produced  has a resolution of
34$^{\prime\prime}$ $\times$ 16$^{\prime\prime}$ at a
position angle of $-$174${\degr}$, and an rms noise of 0.5 mJy~beam$^{-1}$,
representing almost an order of magnitude improvement 
in sensitivity over the image of \citet{dub91}.

\section{Results}
\subsection {The new VLA image of \object{Puppis A}}
\begin{figure*}
\centering
\includegraphics[width=17 cm]{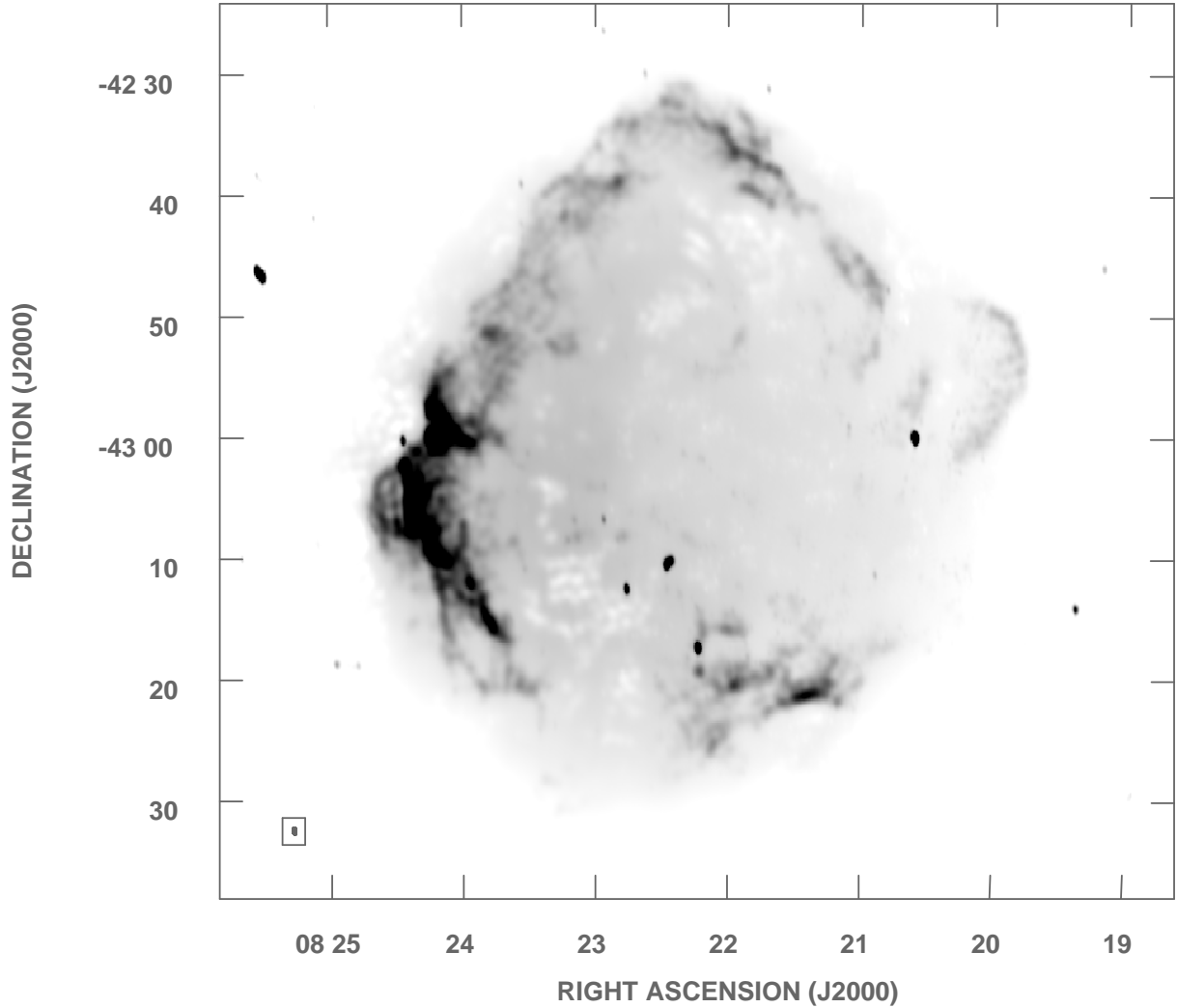}
\caption{The SNR \object{Puppis A} at 1425 MHz. This image was obtained
from a combination of 39 different pointings observed with the VLA in the
DnC and CnB configurations in 2004. Single dish observations from the Parkes Southern Galactic Plane Survey \citep{mcc01} have been added to the data. The angular resolution is
34$^{\prime\prime}$ $\times$ 16$^{\prime\prime}$ (beam shown at lower left),
P.A.=$-$174${\degr}$ with an rms of  0.5 mJy~beam$^{-1}$. The brightness range covered by the grayscale is between 3 and 35
mJy~beam$^{-1}$.
\label{1425}}
\end{figure*}

\begin{figure*}
\centering
\includegraphics[width=17cm]{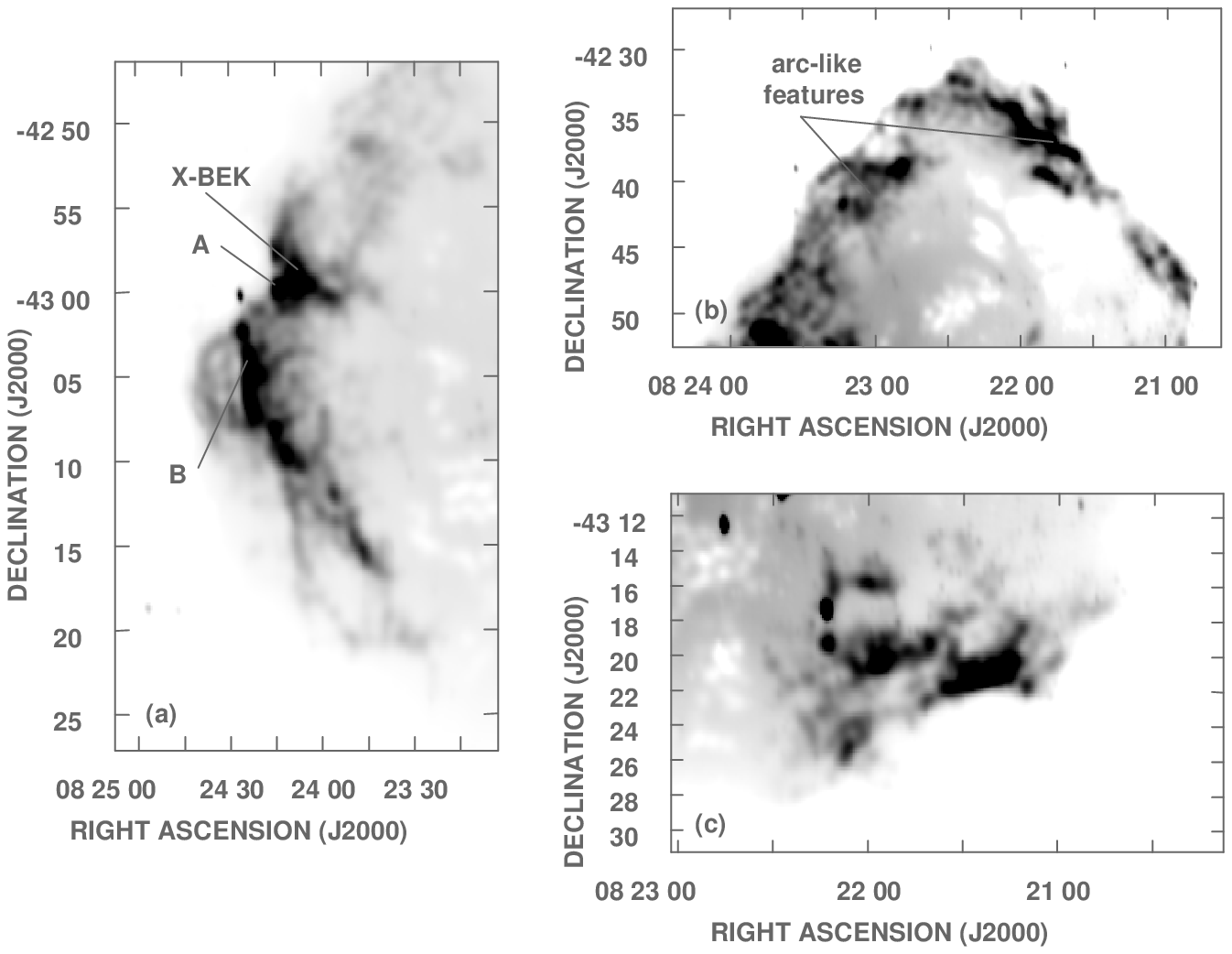}
\caption{Enlargements of the new VLA 1425 MHz image.
\bf (a) \rm E region, containing the brightest features of the remnant.
Feature A marks the position of a bright point-like emission region,
while B indicates the location of the straight N-S filament.
The grayscale ranges from 5 to 52 mJy~beam$^{-1}$.
\bf (b) \rm N part of \object{Puppis A}; the irregular outer border is
prominent. Note the presence of ``wave-like'' features almost
perpendicular
to the shock front on the NE and parallel to it on the NW.
The grayscale ranges from 6 to 16 mJy~beam$^{-1}$. \bf (c) \rm
The SW region showing the presence of the only bright feature located at the
outer boundary of \object{Puppis A}. The grayscale varies between 5 and 18
mJy~beam$^{-1}$.
\label{tres-partes}}
\end{figure*}

Fig.~\ref{1425} shows the new VLA 1425 MHz image of \object{Puppis A}.
At this frequency the remnant has the appearance of an asymmetric shell 
with an average diameter of $\sim\,$55$^{\prime}$ or 35 pc at a
distance of 2.2 kpc.
The intensity of the brightest features along the periphery is 
spatially variable,
ranging from $\sim\,$8 mJy~beam$^{-1}$ in the N to  
$\sim\,$60 mJy~beam$^{-1}$ towards the E.
No sharp boundaries are observed, 
although the presence of a
weak radio halo at a mean level of 3.5 mJy~beam$^{-1}$ 
is apparent in the original images extending about 5$^{\prime}$ ahead
of the bright rim.
With the addition of short spacing contributions, 
all the flux density is represented in the image in Fig.~\ref{1425}. This
image shows
diffuse interior emission at an average level of 8 mJy~beam$^{-1}$.

These new sensitive radio observations
show no evidence for any radio counterpart 
($3\,\sigma$ of 1.5 mJy~beam$^{-1}$) 
to the X-ray point source 
\object{RX J0822$-$4300} nor for a pulsar wind nebula, 
consistent with
previous negative results obtained by \citet{gae00}.

Enlargements showing the detailed 1425 MHz total intensity morphology in
three selected regions of the \object{Puppis A} SNR are shown in 
Fig.~\ref{tres-partes}.
Due to the high dynamic range ($\approx$ 1000:1) of the 1425 MHz image, several
grayscale ranges are used to display the different relevant regions.
Fig.~\ref{tres-partes}a shows the E portion of the SNR
shell, Fig.~\ref{tres-partes}b displays the N part 
of \object{Puppis A},
while Fig.~\ref{tres-partes}c shows the locations of the 
brightest 
features in the SW edge of the remnant.
The new high resolution image reveals that the boundaries 
of \object{Puppis A} 
are quite structured, in contrast with the 
smooth
interior. The outer border consists of a series of 
arc-like features, resembling a ``wave-like'' structure.
Such features are more evident in the 
N half of \object{Puppis A} (Fig.~\ref{tres-partes}b),
although some are detected to the SW. 
We note that along the NW and SW borders, the ``wave-like''
features are observed tangential to the 
outer rim, but their orientation 
changes to approximately radial (i.e perpendicular to
the local shock front) along the NE border.

\begin{figure*}
\centering
\includegraphics[]{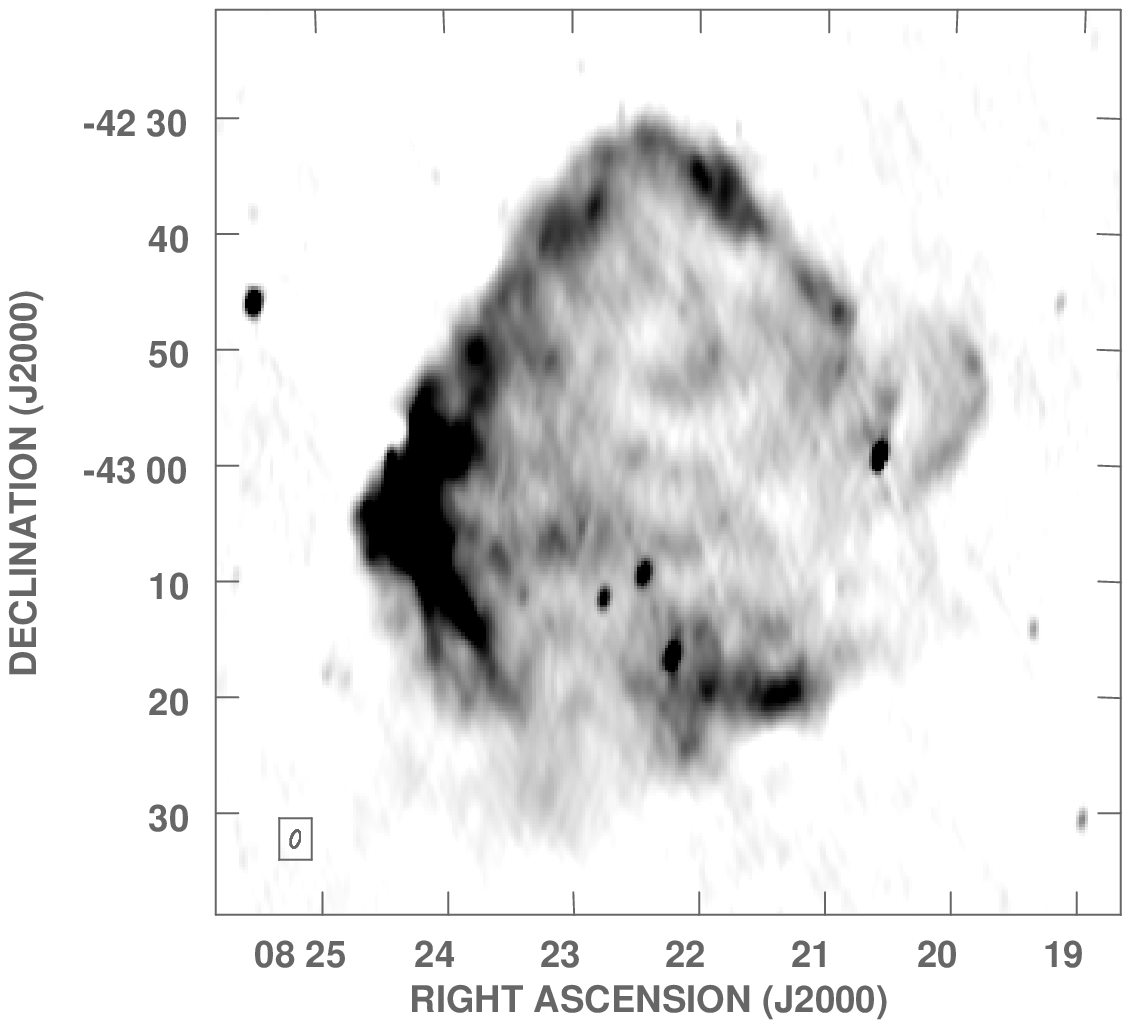}
\caption{Radio continuum image of SNR \object{Puppis A} at 327 MHz obtained
with the VLA in the CnB configuration in 1988 and now re-processed.
The interferometric data were combined with single dish data
from MPIfR Bonn radiotelescope \citep[at 408 MHz,][]{has81}.
The final beam size
is 90$^{\prime\prime}$ $\times$ 45$^{\prime\prime}$ (shown at lower left) at
a position angle of 164${\degr}$,
and the rms noise level is 10 mJy~beam$^{-1}$.
The grayscale ranges from 35 to 350  mJy~beam$^{-1}$.
\label{327}}
\end{figure*}

\citet{pet82} reported the detection of bright X-ray emission near
R.A.$= 08^{\mathrm{h}}\, 24^{\mathrm{m}}$, Decl.$= -43^{\circ}\, 00^{\prime}$
(J2000), 
the BEK
(see Fig.~2a). Intense radio emission is observed coincident
with the X-ray emitting area (see Fig.~\ref{xray} below). 
In addition, the new high resolution 1425 MHz observations 
reveal the presence of a 
compact (size $\sim\,$45$^{\prime\prime}$)
feature of enhanced synchrotron emission (feature A in
Fig.~2a), exactly at the site
where the shock front forms an angle, near
R.A.$= 08^{\mathrm{h}}\, 24^{\mathrm{m}}\, 18^{\mathrm{s}}$,
Decl.$= -43{\degr}\, 00^{\prime}\, 00^{\prime\prime}$ (observed in
Fig.~2a as a 
bright region adjacent to the BEK).
To the S of this feature, 
the presence of a
$\sim\,$6$^{\prime}$ prominent, straight filament (labeled as B in
Fig.~2a) is noticed
near 
R.A.$= 08^{\mathrm{h}}\, 24^{\mathrm{m}}\, 22^{\mathrm{s}}$, running 
N-S
from Decl.$= -43{\degr}\, 02^{\prime}$ to Decl.$= -43{\degr}\, 08^{\prime}$.
To the E of this filament, a fainter semi-circular
protrusion (extending about 6$^{\prime}$) is observed. 
The SW region of \object{Puppis A} (Fig.~2c) shows the only bright feature 
located on the edge 
of the SNR. Short arcs (size $\sim\,$4$^{\prime}$), similar to those observed 
to the NW, are
also observed in this region, although in this case they appear displaced 
by $\sim\,$8$^{\prime}$ from the limb to the interior.

Fig.~\ref{327} shows the re-processed image at 327 MHz of Puppis A.
These data were used with the 1425 MHz image for the spectral analysis
discussed in Section 4.
\section{The spectral properties}
\subsection{The integrated spectrum}
We have calculated integrated flux densities
for \object{Puppis A} based on the new images, obtaining:
S$_{\mathrm{327\,MHz}}\,=\,$263 $\pm$ 20 Jy and
S$_{\mathrm{1425\,MHz}}\,=\,$114 $\pm$ 8 Jy. The errors reflect the 
uncertainties both in the choice of background emission and integration 
boundaries, in addition to the inherent rms noise of the images. 

\begin{figure}
\resizebox{\hsize}{!}{\includegraphics{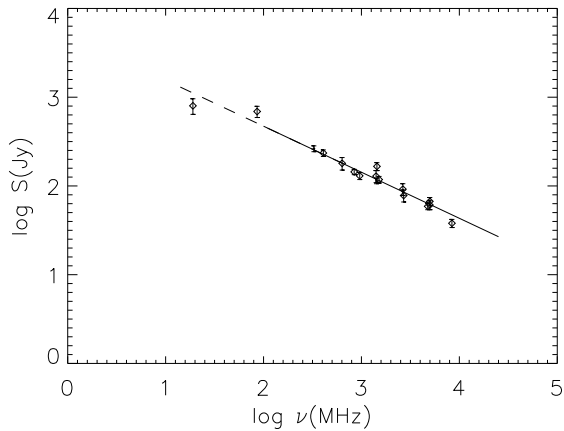}}
\caption{Radio spectrum of \object{Puppis A}
from the scaled flux density values listed in Table~1.
The data from this paper are shown by filled circles (327 and 1425 MHz). The
linear fit to the values of flux density indicates a global spectral 
index of
$-$0.52 $\pm$ 0.03. The dashed line shows the linear 
fit extrapolated to 19 and 86 MHz, at which frequencies no correction
has been applied to bring the measured values to the \citet{baa77} scale.
\label{spectrum}}
\end{figure}

To construct accurate spectra of individual SNRs it is necessary that
all flux density values be brought on to the same absolute flux
density scale.
Table~1 contains the flux densities  
between 19 and 8400 MHz. All the flux densities 
for frequencies above 400 MHz are on the \citet{baa77} scale.
For those below that frequency no scaling was applied.
Fig.~\ref{spectrum} shows the resulting radio 
spectrum of the \object{Puppis A} SNR; the new determinations are 
plotted as filled circles.
The flux density values are adequately fitted by a single power law 
of slope $\alpha\,=\,$$-$0.52 $\pm$ 0.03, 
in good agreement with 
estimates presented by \citet{mil69} ($\alpha$ $\simeq$$-$0.50) and
\citet{dub91} ($\alpha$ $\simeq$$-$0.53). 

\subsection{The spatial spectral distribution}
It is important to discern whether the electron energy spectrum is a global
property of this SNR, or if  
it is locally affected by the neutron star and/or
the shock-ISM interaction. Hence,  
we performed a careful study of the variations in the spectrum as a 
function of position within the remnant.

\begin{table}
\label{fluxes}
\caption{Integrated flux densities for the SNR \object{Puppis A}}
\centering
\begin{tabular}{lll}  \hline \hline
Frequency &  Scaled flux  & References  \\
   (MHz)  &  density (Jy)    & \\ \hline
19        &   800 $\pm$ 160$^{{\mathrm(a)}}$ & \citet{ris58} \\
86        &   690 $\pm$ 100$^{{\mathrm(a)}}$ & \citet{mil60} \\
327       &   263 $\pm$ 20  & This work$^{{\mathrm(b)}}$ \\
408       &   235 $\pm$ 20   & \citet{has81} \\
635       &   180 $\pm$ 29 & \citet{mil69} \\
843       &   144 $\pm$ 10 & \citet{are90} \\
960       &   130 $\pm$ 12  & \citet{har62} \\
1410      &   129 $\pm$ 20 & \citet{mil69} \\
1425      &   114 $\pm$ 8 & This work$^{{\mathrm(b)}}$ \\
1440      &   166 $\pm$ 17  & \citet{mat62} \\
1515      &   118 $\pm$ 10 & \citet{dub91} \\
2650      &   92 $\pm$ 14  & \citet{mil69} \\
2700      &   78 $\pm$ 12 & \citet{mil71} \\
4750      &   59 $\pm$ 5 & \citet{mil93} \\
5000      &   67 $\pm$ 7 & \citet{mil69} \\
5000      &   61 $\pm$ 7 & \citet{mil71} \\
8400      &   38 $\pm$ 4 & \citet{mil93} \\ \hline
\end{tabular}
\begin{list}{}{}
\item[$^{\mathrm{(a)}}$] No correction to \citet{baa77} scale was applied.
\item[$^{\mathrm{(b)}}$] Flux density scale from VLA Calibrator Manual,
http:/www.aoc.nrao.edu/$\sim$gtaylor/calib.html.
\end{list}
\end{table}

The previous study of the spectral index distribution was
carried out by \citet{dub91} using VLA observations at 327 and 
1515 MHz.
The distribution of the spectral indices was estimated by integrating flux
densities 
within five rectangular boxes (of different sizes according to the region, 
varying between approximately 16 and 190 arcmin$^{2}$), and 
comprising the brightest portions of \object{Puppis A}. 
An index as steep as $\alpha\,$$\sim\,-0.7$ was suggested for the E border,
while for the other four regions the spectral index was found to be 
about $-$0.5.
Using more recent spectral analysis techniques and the new VLA 1425 MHz data, 
we are able to produce a 
detailed analysis of the spatial spectral transition from the bright
features to the smooth weaker interior of this SNR.

To accurately determine the spectral index distribution 
from the comparison of two interferometric images at different 
frequencies, it is necessary to ensure that the corresponding images are 
matched in \it uv \rm coverage.  
As mentioned in Section 2.2, the data at
1425 MHz were processed by applying appropiate weights to the visibilities
to obtain an image with the same beam size and beam shape as the
327 MHz image. Also, in order to avoid any positional offsets, the images 
were aligned and interpolated to identical projections (field center, pixel 
separation, etc). In addition, we checked for
possible flux density calibration
bias. That is, an extrapolation
back to zero flux density at one frequency should correspond to zero flux
density at the other frequency, otherwise there is a
zero-point offset that must be corrected.
We have checked the 327 and 1425 MHz images, finding only an
insignificant offset of 0.1 mJy~beam$^{-1}$ between
frequency pairs. Therefore, no further correction was applied.

Based on these images, we carried out the spatial spectral analysis. 
This was
done using two different methods in order to confirm that the observed
spectral features are real, independent of any possible
contamination from the databases used for the comparison. 

\begin{figure}
\includegraphics[width=9 cm]{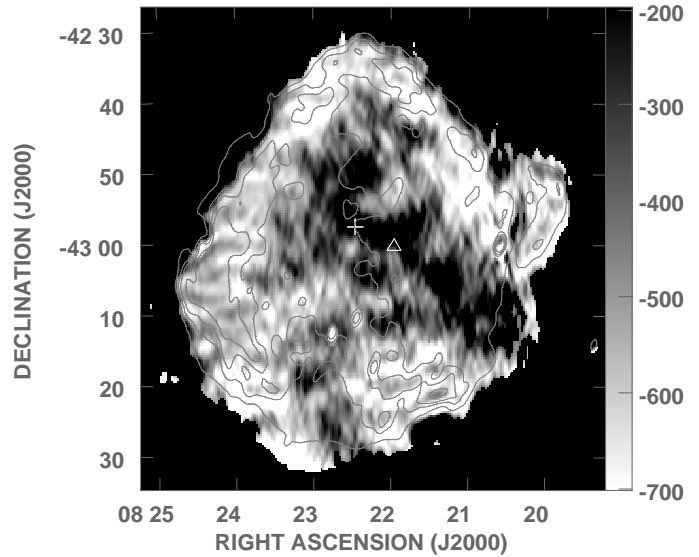}
\caption{Grayscale image of the spectral index distribution across the
SNR \object{Puppis A} as derived between 327 and 1425 MHz. The wedge displays 
the spectral index value ($\times 1000$).
The 1425 MHz intensity contours
plotted for reference are 36, 60, 90 and 160 mJy~beam$^{-1}$.
A cutoff at 5$\,\sigma$ was applied at both frequencies.
The plus sign indicates the
geometrical center of the SN explosion, while the triangle marks the 
position of the neutron star \object{RX J0822$-$4300}. 
\label{alphamap}}
\end{figure}

In Fig.~\ref{alphamap} we show the spectral index 
distribution of \object{Puppis A} as obtained from the direct comparison of
the matched images at 327 and 1425 MHz 
after clipping at the 5$\,\sigma$ level for both
frequencies. 
We find that the spectral index varies across the SNR approximately between
$\alpha \sim\,$$-$0.8 and $-$0.2.  
In this grayscale representation, darker means
flatter spectrum. 
It is interesting to note that the feature with the flattest spectrum 
(seen as a black band running from NE to SW in the interior of 
\object{Puppis A})
includes the location of the neutron star 
\object{RX J0822$-$4300} and 
of the geometrical center of the explosion 
as derived from optical observations by \citet{win88} 
(shown in Fig.~\ref{alphamap} as a triangle and a plus sign
respectively). 
These positions are known with an accuracy of tenths of an arcsec for
\object{RX J0822$-$4300} and $\sim\,$1 arcmin  
for the proposed center of the explosion.
We can confidently assume that this spectral feature is real since
both images have recovered all flux density.

The present spectral map
shows that along most of the periphery the spectrum is 
clearly steep (steeper than $\alpha$=$-$0.7). 
To the W, the ear-like feature shows a combination of different 
structures with variable spectral indices. 

Curiously the spectral pattern observed near the bright eastern radio
features (approximately near R.A.$\sim 08^{\mathrm{h}}\, 24^{\mathrm{m}}$,
from Decl. $\sim -42^{\circ}\, 55^{\prime}$ to $\sim -43^{\circ}\, 
10^{\prime}$) does not correlate with the morphology observed in total power
(see Fig.~\ref{1425}).
Although it is somewhat flatter than the rest of the shell, it is apparent 
that the spectrum varies following a pattern
of parallel fringes that alternate steeper and flatter spectral indices. 

In order to test the spectral index distribution determined above, we 
also
constructed a gallery of tomographic maps.
Spatial tomography between two images is a technique developed 
by \citet{kat97}, based on the use of a test spectral index 
$\alpha_{\mathrm{t}}$. A tomographic map is calculated by scaling the 
brightness
in the higher frequency image (S$_{\mathrm{1425\,MHz}}$, in our case) by the
test spectral index, and subtracting this scaled image from the lower 
frequency image (S$_{\mathrm{327\,MHz}}$). Then, a tomographic gallery is 
obtained by calculating 
S$_{\mathrm{t}}\,=\,$S$_{\mathrm{327\,MHz}}\,-\,\left(\frac{\mathrm{327}}{\mathrm{1425}}\right)^{\alpha_{\mathrm{t}}}\:$S$_{\mathrm{1425\,MHz}}$, 
stepping 
through a range of spectral indices $\alpha_{\mathrm{t}}$.  
Features which have a spectral index
identical to the test value will vanish in the tomographic map. Spatial
components that have different spectral indices will appear as  positive
or negative features depending upon whether the spectrum is steeper or 
flatter than the assumed test value. 
This method has been succesfully applied to analyze spatial 
spectral variations in the SNRs of \object{Tycho} \citep{kat00}, 
\object{Kepler} \citep{del02},
\object{G292.0+1.8} \citep{gae03} and in the study of the radio galaxies 
\object{3C\,67}, \object{3C\,190} 
and \object{3C\,449} \citep{kat97, kat97b}. The technique is
particularly useful to disentagle components with different spectra 
that can overlap along the line of sight. Also, when the goal is to locate 
small-scale spectral variations, this method can provide a more 
accurate 
picture than the simple comparison between images at different frequencies. 

\begin{figure*}
\centering
  \includegraphics[width=11 cm]
{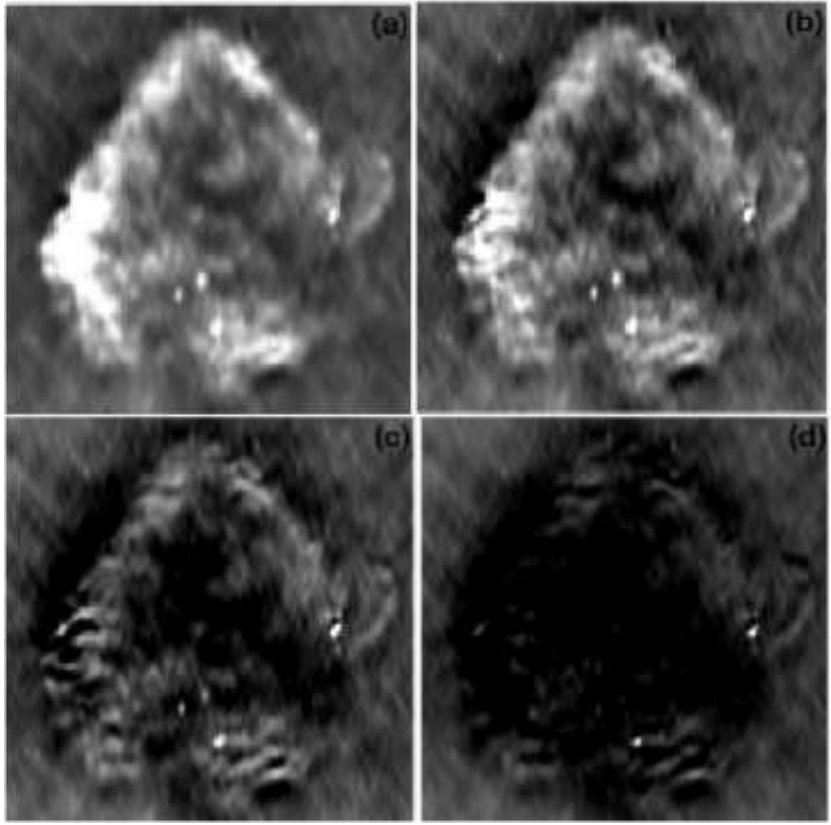}
\caption{Series of tomographic images for the SNR \object{Puppis A}
constructed from data at 327 and 1425 MHz,
which were matched in uv coverage.
\bf (a) \rm $\alpha_{\mathrm{t}}$=$-$0.2;
\bf (b) \rm $\alpha_{\mathrm{t}}$=$-$0.4; \bf (c) \rm 
$\alpha_{\mathrm{t}}$=$-$0.6; and 
\bf (d) \rm $\alpha_{\mathrm{t}}$=$-$0.8.
Bright areas have radio
spectral indices steeper than the assumed $\alpha_{\mathrm{t}}$ and dark areas
imply flatter spectral indices.
The same grayscale is used to display the four
images.
\label{tomo}}
\end{figure*}

In Fig.~\ref{tomo} we show the resulting tomographic
images for four test values:
$\alpha_{\mathrm{t}}$=$-$0.2 (Fig.~\ref{tomo}a), $-$0.4 (Fig.~\ref{tomo}b), 
$-$0.6 (Fig.~\ref{tomo}c), and $-$0.8 (Fig.~\ref{tomo}d).
In these maps the spatial components appear as
positive (light) or negative (dark) 
features for a spectrum steeper or flatter than 
the assumed test value respectively. 
The point sources overlapping the remnant are always seen
as positive residuals in the tomographic images (white spots) as they
have spectral indices steeper than $-$0.8, thus supporting their 
extragalactic nature,  
as already noted by \citet{mil83} and \citet{dub91}.

The most interesting tomographic images are those obtained from the 
comparison 
with $\alpha_{\mathrm{t}}$=$-$0.4 (Fig.~\ref{tomo}b) 
and with $\alpha_{\mathrm{t}}$=$-$0.6 (Fig.~\ref{tomo}c), where,
as noticed before, a
notable striation is evident along the periphery. 
It is worth noting that this spectral pattern formed by short
fringes with $\alpha$ alternatively steeper and flatter than the
background, mimics the ``wave-like'' morphology noticed in the total 
power image along the NE, NW and S borders. That is, the spectral components
appear almost 
parallel to the outer boundary to the NW and SW, and transverse 
to it along the E side.

The newly identified point-like feature to the E of \object{Puppis A} 
(feature A in Fig.~\ref{tres-partes}a), has a spectrum steeper 
than $-$0.8
(i.e. it appears as a positive residual in all tomographic 
images). 
We thus conclude that, in all probability, it is an extragalactic 
object.

The central band seen in Fig.~\ref{alphamap} disappears against the
local background in the tomographic image traced with 
$\alpha_{\mathrm{t}}$=$-$0.2 (Fig.~\ref{tomo}a), thus confirming its very flat
spectrum.

The grayscale corresponding to the 
$\alpha_{\mathrm{t}}$=$-$0.8 image (Fig.~\ref{tomo}d) 
reveals that the spectral distribution in \object{Puppis A} has practically 
no components steeper than this value. 

\section{Comparison between radio and X-ray emission}
The X-ray properties of \object{Puppis A} have been investigated 
using the \it Einstein\rm, \it ASCA \rm and \it ROSAT \rm
telescopes \citep[][ and references therein]{pav99}.
The highest resolution X-ray image of the entire source was obtained 
with \it ROSAT \rm (HRI) in the energy band from 0.1 to 2 keV  \citep{pet96}.

\begin{figure*}
\centering
\includegraphics[]{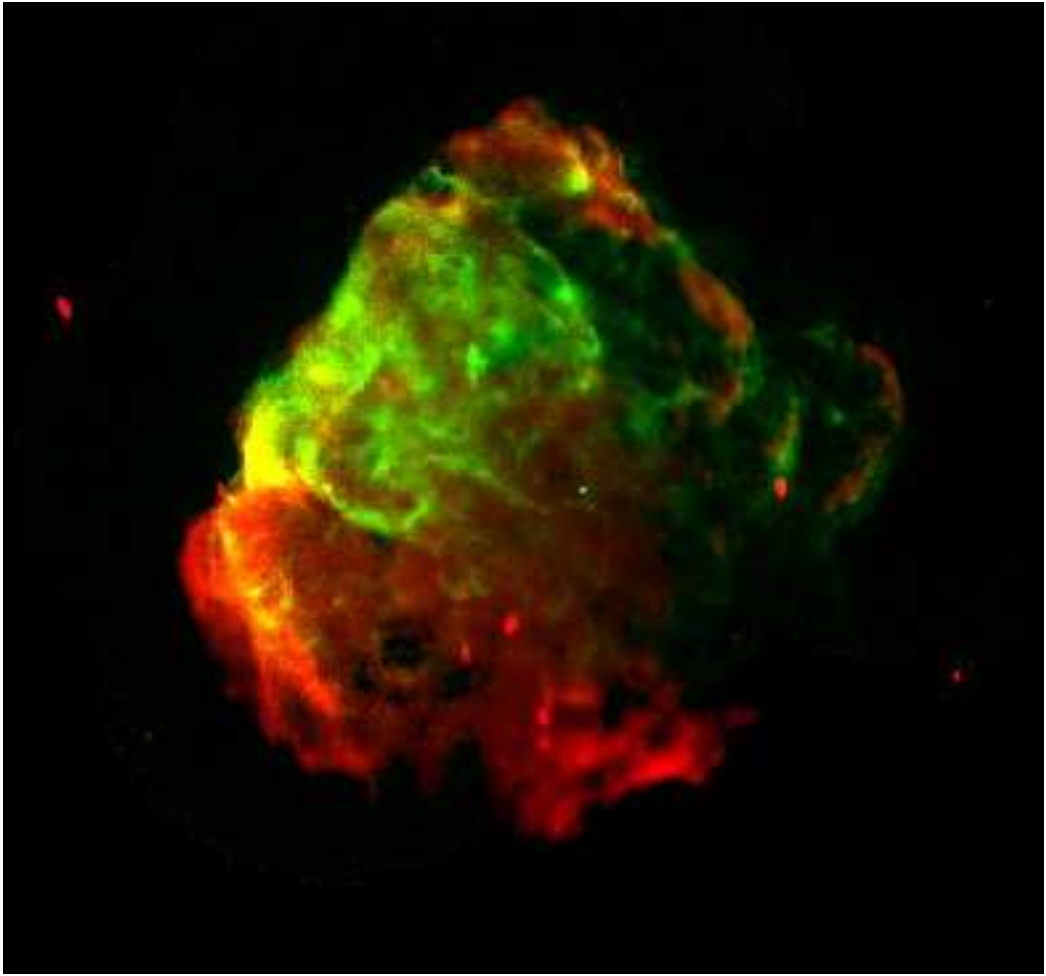}
\caption{A high resolution X-ray/radio comparison of the
\object{Puppis A} SNR.
The green image corresponds to soft X-ray emission from
\it ROSAT \rm in the 0.1-2 keV range \citep{pet96}, while in red the
1425 MHz radio emission is shown. The yellow regions are areas where emission
in both spectral ranges overlap. The X-ray image is smoothed to
the resolution of the radio image, 34$^{\prime\prime}$ $\times$ 16$^{\prime\prime}$,
P.A.=$-$174${\degr}$.
\label{xray}}
\end{figure*}

In Fig.~\ref{xray} we present a false color image showing the spatial 
comparison  
between the radio image at 1425 MHz (in red) and the soft X-ray 
emission 
(in green) for the entire 
remnant. For the first time a comparison with a spatial 
resolution $<$ 30$^{\prime\prime}$ is possible.
Features where both spectral bands overlap are shown in yellow.

The X-ray emission associated with \object{Puppis A}  has been 
interpreted as resulting from several thermal components \citep{ber94}.
The N half of \object{Puppis A} is clearly the brighter and more
complex in X-rays. A network of small
scale structures that include filaments, short arcs and knots are, in
general, poorly correlated with radio features. The brightest features 
are preferentially located to the E although the N-NW rim is also well 
delineated by bright X-ray filaments.
Fig.~\ref{xray} demonstrates that the most striking radio/X-ray 
correspondence
occurs predominantly along the 
E limb of the remnant.
Three features are well matched: 
the BEK, the curved filament near
R.A.$=\, 08^{\mathrm{h}}\, 23^{\mathrm{m}}\, 20^{\mathrm{s}}$,
Decl.$=\,-43{\degr}\, 02^{\prime}$ (see below) and the indentation around
R.A.$\sim 08^{\mathrm{h}}\, 24^{\mathrm{m}}\, 30^{\mathrm{s}}$,
Decl.$\sim -43{\degr}\, 00^{\prime}$.
In contrast, little  radio/X-ray agreement is observed in the
SE sector of \object{Puppis A};
the radio emission extends over 5$^{\prime}$ 
beyond the 
X-ray limb. Moreover, the protrusion ahead of the bright,
straight 
N-S radio filament (feature B in Fig.~\ref{tres-partes}a) 
lacks a counterpart in X-rays.
A similar radio/X-ray comparison cannot be carried out to the SW of 
\object{Puppis A}
due to a missing \it ROSAT \rm pointing in this direction
(see the \it ROSAT \rm archive); thus the reconstruction of the image 
is likely incomplete in this region.                                                                                
Towards the NW border of \object{Puppis A} 
the X-ray emission accompanies the ``wave-like'' morphology 
revealed in the radio, 
with the same orientation parallel to the shock front. The 
dense X-ray knot located 
immediately behind the NW shock front
(the ``bright northern knot'', BNK,
R.A.$ \sim \, 08^{\mathrm{h}}\, 22^{\mathrm{m}}$,
Decl.$\sim \, -42\,{\degr}\, 37^{\prime}\,$) is prominent in both spectral 
domains.
\citet{pet82} have suggested that this knot originated as a result
of the interaction with a dense neutral 
cloudlet, similiar to  the BEK.
In the radio domain, the  spectrum of this knot between 327 and 1425 
MHz 
has a spectral index $\alpha\,$$\sim-\,$0.5 (see Section 4.2.2), similar to 
the global spectral index of this SNR.
A careful spectral study of this clump in X-rays would 
help towards understanding its origin.

\begin{figure*}
\centering
\includegraphics[width=17cm]{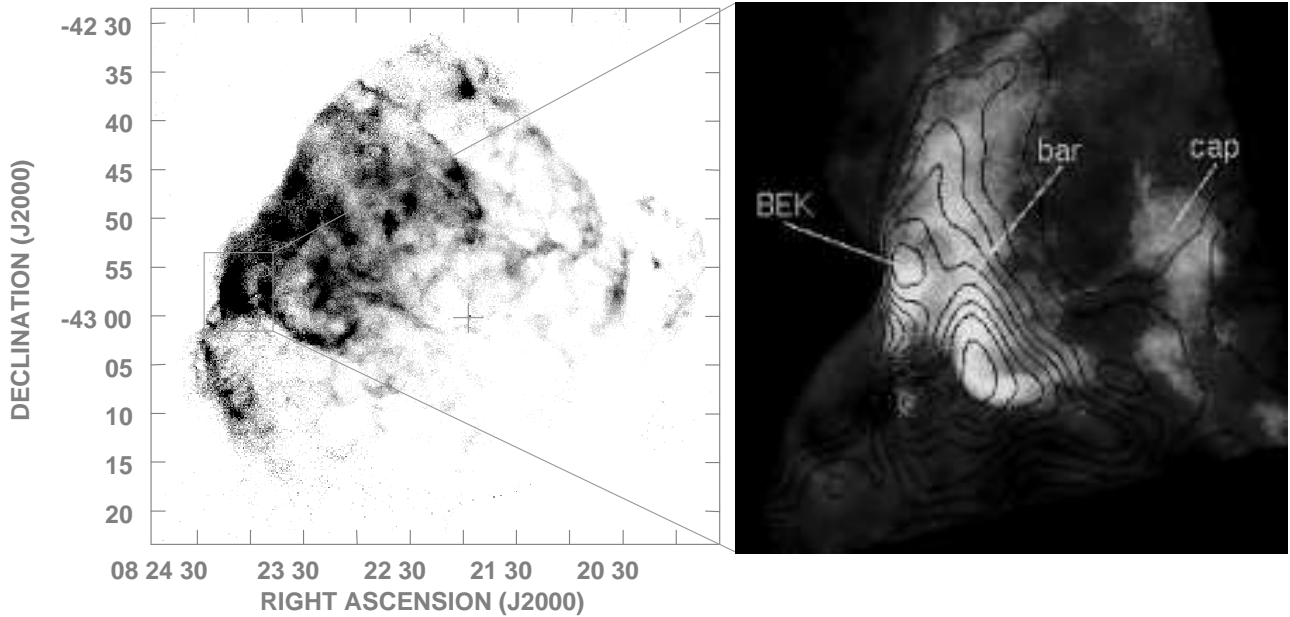}
\caption{
\it (Left) \rm \it ROSAT \rm HRI mosaic of the \object{Puppis A} SNR.
The box indicates
the region of the 
E limb observed with \it Chandra \rm \citep{hwa05}.
The plus (+) symbol shows the position of the compact source
\object{RX J0822$-$4300}.
\it (Right) \rm The grayscale representation combines soft
($\sim\,$0.4-0.7 keV), medium ($\sim\,$0.7-1.2 keV), and hard band
($>$1.2 keV) images around the bright eastern knot, while the superposed
contours trace the radio synchrotron emission at 1425 MHz at 18, 23, 29, 36, 43, 50, 56, 61, 68, 74, 80, and 87 mJy beam$^{-1}$.
\label{una}}
\end{figure*}

\citet{hwa05} have used the \it Chandra \rm X-ray telescope to 
resolve the fine structure
around the BEK over a field of view of 8$^{\prime}$ $\times$ 8$^{\prime}$. 
These authors describe the inner structure in this part of the
remnant as consisting of: (1) a bright compact knot,
(2) a curved vertical feature (named the ``bar'') 
about 1$^{\prime}$ 
W behind the shock,
and (3) a smaller bright cloud (the ``cap'') separated from the 
``bar'' by faint
diffuse emission. Nomenclature for these X-ray features is 
contained in Fig.~\ref{una}.

To study the relation between the X-rays and
the radio components in the \object{Puppis A} SNR,
in Fig.~\ref{una} we show a direct comparison of the new 1425 MHz VLA
image with the \it Chandra \rm X-ray image obtained from the 
combination
of the soft, medium, and hard X-rays (note that the \it Chandra \rm 
image has an angular resolution about 5 times finer than the radio 
image).
Fig.~\ref{una} reveals that the X-ray emitting plasma does not 
coincide with radio features at a detailed level. 
However, on a larger scale reasonable agreement is observed.
The radio emission is  compressed in the
indented 
eastern X-ray boundary, confirming that the shock wave is wrapping 
around an obstacle in the neutral ISM as was earlier proposed by 
\citet{hwa05}.
A local radio maximum matches the BEK, as shown in Fig.~\ref{una}. 
In this portion of the SNR, this feature
is not the  brightest
feature in radio as is the case for X-rays. 
The most intense radio
emission is found to the E of the X-ray ``bar''
(around R.A.$=\, 08^{\mathrm{h}}\, 24^{\mathrm{m}}\, 14^{\mathrm{s}}$,
Decl.$=\, -42{\degr}\, 59^{\prime}\, 49^{\prime\prime}$), where it 
correlates with a secondary X-ray peak (see Fig.~\ref{una}).
No conspicuous radio feature can be associated with
the ``cap''.
The region of faint diffuse X-ray emission between
the ``bar'' and the ``cap'' is a region of low surface brightness
radio emission.

In summary, based on the present comparison 
only moderate agreement between radio and
X-ray emission has been observed at a detailed level. 
This conclusion is not surprising since the X-rays can be associated
with thermal radiation from a hot plasma, while the radio continuum
arises via synchrotron emission from relativistic electrons.

\section {Is Puppis A a new case of a Galactic SNR shaped by energetic jets?}
\begin{figure*}
\centering
\includegraphics[width=17cm]{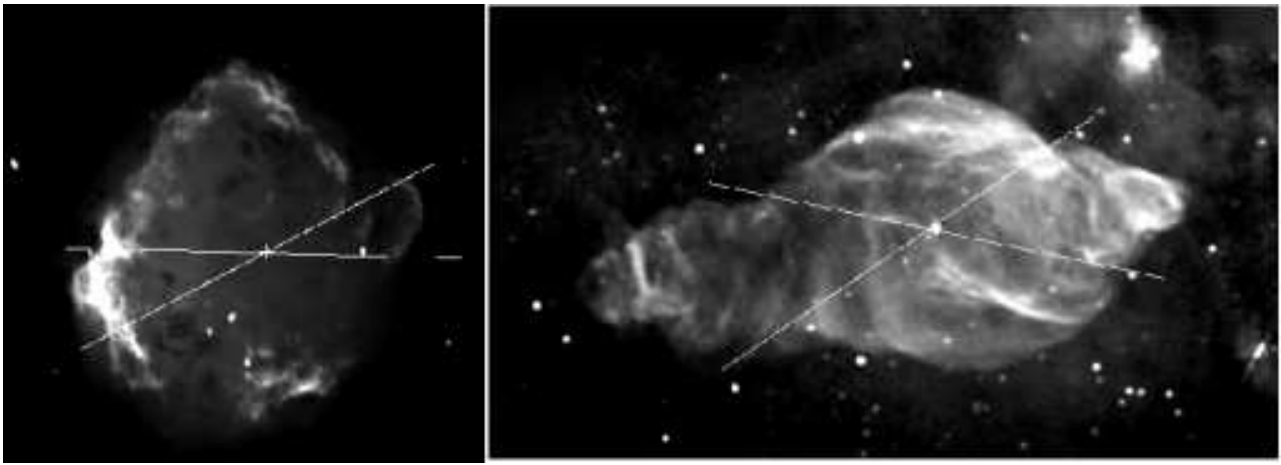}
\caption{Comparison of the SNR \object{Puppis A} \it (left) \rm with the SNR
\object{W50} that contains the central X-ray
binary \object{SS433} \citep[\it right\rm, from][]{dub98}.
In \object{Puppis A}, the
white cross shows the location of the X-ray point source
\object{RX J0822-4300},
a radio-quiet neutron star proposed to be the compact remnant of
\object{Puppis A}
explosion. The white intersecting lines delimit the ``cones''
proposed to be associated with \object{Puppis A} in analogy with the case of
\object{W50}. In the image of the SNR \object{W50}, the white lines trace the
volume defined by the oppositely directed, precessing, 
relativistic jets.
\label{jets}}
\end{figure*}

As noticed before, the general shell-like morphology of \object{Puppis A} 
appears
distorted to the E by the protrusion near 
R.A.$\sim 08^{\mathrm{h}}\, 24^{\mathrm{m}}\, 30^{\mathrm{s}}$, 
Decl.$\sim -43{\degr}\, 06^{\prime}$ and to the W by the ``ear'' around 
R.A.$\sim 08^{\mathrm{h}}\, 20^{\mathrm{m}}\, 00^{\mathrm{s}}$, 
Decl.$\sim -42{\degr}\, 54^{\prime}$ (Fig.~\ref{1425}, Fig.~2a and 
Fig.~\ref{xray}). If we
trace two intersecting lines oriented about a 
position angle of $\sim\,$100${\degr}$ (from N through E),
with a semi-opening angle of $\sim\,$15${\degr}$, crossing at the position 
of the neutron star candidate \object{RX J0822-4300}, the appearance of 
\object{Puppis A}
strikingly resembles that of the SNR \object{W50}. The shell of 
\object{W50} is distorted by the precessing 
jets that originate in the central compact source \object{SS433}, and 
Fig.~\ref{jets} illustrates the similarity between the two 
SNRs. \rm In 
Fig.~\ref{jets} ({\it left}) we have plotted a sketch
of the  hypothetical ``cones'' overlapping the new radio image of
\object{Puppis A}, while
on the right side an image of the 
SNR \object{W50} at 1.4 GHz \citep[from ][]{dub98} is shown 
with the precession cones of $\sim\,$20${\degr}$ semi-opening angle 
delimited  by the intersecting white lines. In the case of \object{W50}, 
the elongated shape has been
interpreted as being due to the action of the two oppositely directed
relativistic jets injected by the X-ray binary \object{SS433}.
The present comparison shows that not only the general
appearance with symmetrical lateral extension is
similar for both SNRs, but that the analogy also extends to
particular features. For example, the bright straight N-S filament in
\object{Puppis A} (named B in Fig.~2a) looks very similar to the 
straight radio filament that stands out near the E 
extremity of \object{W50}, and which is
thought to be the radio ``hot-spot''
produced by the interaction of the relativistic jet from 
\object{SS433} with 
the ISM \citep{saf97}. Beyond these bright, straight filaments,
diffuse radio emission is observed in both cases.

The compact X-ray source \object{RXJ 0822-4300} in 
\object{Puppis A} has been proposed to be the
compact stellar remnant formed in the SN explosion \citep{pav99}.
It is a radio-quiet neutron star belonging to the class of ``Central
Compact Objects'' (CCOs) in SNRs whose nature remains unclear 
\citep{pav02}. Based on detailed HI observations
carried out in the direction of \object{RXJ 0822-4300}, \citet{rey03}
suggested that this CCO could be ejecting two oppositely directed  
jets that are
creating a bi-lobed cavity in the HI. This elongated hole is about
15$^\prime$
long, oriented in the NE-SW direction, with a position angle of about
60${\degr}$. Both the X-ray point source and the
position of the geometrical center
of the explosion \citep[suggested to be at  
R.A.$\sim 08^{\mathrm{h}}\, 22^{\mathrm{m}}\, 27^{\mathrm{s}}.5$, 
Decl.$\sim -42{\degr}\, 57^{\prime} \, 29^{\prime\prime}$, ][]{win88} 
lie within the HI hole. Thus, the possible HI signature of jets is
not exactly aligned with the direction of the hypothetical 
cones that connect the large scale E and W extensions. However, it
cannot be discounted
that the HI feature
is related to  the path of the neutron star from the center of explosion
to its present
position, while the large scale cones are likely to be associated with 
the persistent action of
jets on the surrounding SNR shell.

In the X-ray domain, no conspicuous feature is apparent
in the immediate vicinity of \object{RXJ 0822-4300} 
(see Fig.~\ref{xray} and Fig.~\ref{una} {\it left}).
The first
bright X-ray features within the hypothetical cones are the thin
filament to the E near R.A.$\sim 08^{\mathrm{h}}\, 24^{\mathrm{m}}\, 
15^{\mathrm{s}}$, 
Decl.$\sim -43{\degr}\, 05^{\prime}$ and to the west near 
R.A.$\sim 08^{\mathrm{h}}\, 20^{\mathrm{m}}\, 30^{\mathrm{s}}$,
Decl.$\sim -43{\degr}\, 00^{\prime}$ (see Fig.~\ref{una} {\it left} for a 
representation of the X-ray emission
in equatorial coordinates). At the assumed distance
of 2.2 kpc, these features
occur at about $\sim\,$32 pc E and $\sim\,$22 pc W of the central point
X-ray source. In the case of \object{W50}, the X-ray lobes are 
symetrically
displaced E and W of \object{SS433} starting at about 15$^\prime$ from
the center (approximately 17 pc at the distance of \object{W50}) and have 
a knotty structure with enhanced emission at about 35$^\prime$ (i.e. about
40 pc) from \object{SS433} \citep{saf97}. The displacement
of the X-ray lobes in \object{W50}  has been interpreted by \citet{saf97}
as due to the expansion of beams in a low density ISM.

A careful study of the X-ray features in the interior of \object{Puppis A}
is highly needed to understand the interaction of the central
neutron star  with the surrounding SNR and to answer the question
as to whether \object{Puppis A} is another example of a
Galactic SNR
distorted by the action of symmetric energetic jets.

\section{Conclusions}
In the current work we have presented a new image of the SNR 
\object{Puppis A} at 1425 
MHz. This image has revealed with superb detail the presence of short
filaments along most of the distorted periphery of \object{Puppis A}.
In particular, in the northern half of \object{Puppis A} the radio 
emission near the limb resembles a ``wave-like'' 
structure, oriented parallel to the
shock front along the NW but almost perpendicular to it
towards the brighter NE limb.
Such a pattern is replicated in the spatial spectral distribution, 
where fringes with analogous orientation alternate between flat and 
steep spectra. 
We propose that the peripheral filaments are a manifestation of 
Rayleigh-Taylor (R-T)
instabilities that are distorting the interface between the ejecta and the
ambient material. As R-T instabilities develop, they stretch and 
compress the magnetic field \citep{jun96}. This produces magnetic field 
amplification
in the ambient medium close to the interface, which in turn enhances 
the synchrotron emission. 
In our model, the R-T fingers observed to the NE are radial, 
while to the
NW they are tangential suggesting that along this border the 
shock is probably in a more advanced evolutionary stage.
This could be related to the possibility that to the NE the shock 
front is expanding in a denser ISM. 
To confirm this hypothesis, it is important to investigate the magnetic field
orientation in the region. A detailed study of 
polarization vectors at high angular resolution is planned for the 
near future. 

Similar short, curved filaments are also observed in the young SNRs of
\object{Tycho} and \object{Cas A}, both in the radio and in 
X-rays, 
although in these cases a
weaker plateau and a thin brighter outer rim is observed ahead of the 
filaments \citep{rey97,hwa04}. In \object{Puppis A}, a faint plateau can be 
clearly seen around almost
all the periphery, but no sharp outer border, presumably 
associated with the foward shock, is detected in this 
intermediate-age SNR. 

The most conspicuous radio features are situated 
at the E border, at the site where the shock front has engulfed a dense ISM 
cloud.
Interestingly enough, the radio counterpart of the BEK and the
narrow straight filament located just south of it
do not appear as 
spectrally distinct features. If these synchrotron maxima are the result
of the interaction with external clouds, then
we are led to the conclusion that
this condition is not reflected in the radio spectrum. 
The spectral map of \object{Puppis A} does not show diferences 
between the E and the W
sides in spite of the gradient in density in the ISM \citep{pet82}. We 
can therefore 
conclude that in this case the shocks are probably still 
non-radiative and the compression is not very high. This contention is 
supported by the lack of optical filaments in the boundaries and the generally 
poor radio/optical correlation \citep{dub91}. 

Another spectral peculiarity in \object{Puppis A} is the
fact that the spectrum steepens in general from the interior to the periphery.
In the first-order Fermi, non-relativistic, test particle 
limit, stronger shocks yield flatter indices up to the $\alpha$=$-$0.5 
limit.
Therefore, based on this acceleration model, higher emissivity 
regions should
have flatter spectrum, while steep spectra are expected in the more diffuse
emission regions \citep[see for example][]{and93}, exactly 
the opposite of 
what we find in \object{Puppis A}. This result is confirmed through the 
tomographic analysis of the spectrum.
It is interesting
to note that the region of \object{Puppis A} with the flattest 
spectrum is a narrow 
band towards the
interior, like a channel that connects the position of the neutron star
\object{RX J0822$-$4300} with the presumed geometrical center of 
explosion. 
The continuous injection of relativistic particles by the neutron star could 
provide an explanation for this spectral characteristic.

Based on a compilation of integrated flux densities over a wide spectral 
range, we have recalculated the global 
non-thermal radio spectrum of the SNR \object{Puppis A}. 
The fit produces a
spectral index of $\alpha$=$-$0.52 $\pm$ 0.03 for the whole SNR.

We have also used the new radio image to perform the first accurate
radio/X-ray comparison for \object{Puppis A}. In 
X-rays,  
\object{Puppis A} is bright on
the E half, and the emission seen by \it ROSAT \rm between 
0.1 and 2 keV is 
very structured. The new radio image, although much more sensitive than
previous ones, does not show detailed correlation with the thermal plasma
imaged in X-rays, except for the bright E features. 
Even there,
the matching is not accurate when the radio emission is compared with the
high resolution \it Chandra \rm image. The only remarkable coincidence 
revealed by the new results occur along the N-NW boundaries, 
where the ``wave-like'' radio features near the shock
front have their exact counterpart in the X-rays.  
A detailed X-ray study along the NW side, 
including the BNK, would be valuable.
To the S-SW of \object{Puppis A},
we were unable to carry out a similar comparison because a 
\it ROSAT \rm pointing is
missing there.

On the other hand, based on the new observations, we have confirmed 
that 
down to $\sim\,$1.5 mJy~beam$^{-1}$ neither a radio counterpart 
to the neutron star
\object{RX J0822$-$4300}, nor a surrounding pulsar wind 
nebula, is observed.

Finally, based on the radio images and their comparison with the X-ray
emission, a  striking similarity between \object{Puppis A} and the SNR 
\object{W50},
which harbors the Galactic microquasar \object{SS433} in its interior, is
seen. This suggests 
that \object{Puppis A} is a candidate for 
being another SNR shaped by internal jets, probably at an earlier 
evolutionary stage.
Detailed central X-ray images and spectral studies would be of great
help to test this suggestion and to understand the nature of the
central X-ray compact object.

\begin{acknowledgements}
We would like to thank Naomi McClure- Griffiths for providing us with
the 1.4 GHz data from the Parkes Southern Galactic Plane Survey.
We are grateful to the referee for all his/her suggestions which certainly
has helped to make this a better paper. This work has been supported by 
Argentina grants 
ANPCYT-PICT 04-14018,
UBACYT A055/04, ANPCYT-PICT 03-11235 and PIP-CONICET 
6433.
\end{acknowledgements}
\bibliographystyle{aa}  
\bibliography{bib-puppis}
\IfFileExists{\jobname.bbl}{}
{\typeout{}
\typeout{****************************************************}
\typeout{****************************************************}
\typeout{** Please run "bibtex \jobname" to optain}
\typeout{** the bibliography and then re-run LaTeX}
\typeout{** twice to fix the references!}
\typeout{****************************************************}
\typeout{****************************************************}
\typeout{}
}

\end{document}